\documentclass[pra,aps]{revtex4}
\usepackage{graphicx}

\begin{document}

\title{Weakly anisotropic frustrated zigzag spin chain}
\author{D.V.Dmitriev}
 \email{dmitriev@deom.chph.ras.ru}
\author{V.Ya.Krivnov}
\affiliation{Joint Institute of Chemical Physics of RAS, Kosygin
str.4, 119334, Moscow, Russia.}
\date{}

\begin{abstract}
The frustrated spin-1/2 model with weakly anisotropic
ferromagnetic nearest-neighbor and antiferromagnetic
next-nearest-neighbor exchanges is studied with use of variational
mean-field approach, scaling estimates of the infrared
divergencies in the perturbation theory and finite-size
calculations. The ground state phase diagram of this model
contains three phases: the ferromagnetic phase, the commensurate
spin-liquid phase and the incommensurate phase. The non-trivial
behavior of the boundaries between these phases and the character
of the phase transitions in case of weak anisotropy are
determined.

\end{abstract}
\maketitle

\section{Introduction}

The quantum spin chains with nearest-neighbor (NN) $J_1$ and
next-nearest-neighbor (NNN) interactions $J_2$ have been a subject
of numerous studies \cite{review}. The model with both
antiferromagnetic interactions $J_1,J_2>0$ (AF-AF model) is well
studied \cite{Haldane,Tonegawa87,Okamoto,Bursill,Majumdar,White}.
Lately, there has been considerable interest in the study of F-AF
model with the ferromagnetic NN and the antiferromagnetic NNN
interactions ($J_1<0$, $J_2>0$)
\cite{Tonegawa89,Chubukov,KO,Vekua,Lu,Nersesyan,Langari}. One of
the reasons is understanding of intriguing magnetic properties of
a novel class of edge-sharing copper oxides which are described by
the F-AF model
\cite{Mizuno,Drechsler1,Drechsler2,Hase,helimagnetism1,helimagnetism2}.
In particular, these copper oxides show at low temperature a
tendency to the formation of the incommensurate state with
helicoidal magnetic ordering.

The Hamiltonian of the F-AF model is
\begin{equation}
H =
-\sum_{n=1}^{N}(S_{n}^{x}S_{n+1}^{x}+S_{n}^{y}S_{n+1}^{y}+\Delta
_{1}S_{n}^{z}S_{n+1}^{z}) +
J\sum_{n=1}^{N}(S_{n}^{x}S_{n+2}^{x}+S_{n}^{y}S_{n+2}^{y}+\Delta
_{2}S_{n}^{z}S_{n+1}^{z})  \label{H}
\end{equation}
where we put $J_1=-1$ and $J_2=J>0$ and the periodic boundary
conditions are implied.

The isotropic case of this model ($\Delta_1 = \Delta_2 = 1$) is
intensively studied last years \cite{Vekua,Lu,DK06,Cabra,Itoi}. It
is known that the ground state of the isotropic version of the
model (\ref{H}) is ferromagnetic at $0<J<1/4$ and it becomes a
singlet incommensurate state for $J>1/4$ \cite{Schilling,Hamada}.
The phase transition at $J=1/4$ is the second order one.

The model with the anisotropy of exchange interactions is less
studied, especially for the case of the small anisotropy. For
example, the phase diagram of the model (\ref{H}) with
$\Delta_1=\Delta_2$ has been studied in Ref.\cite{Aligia} using
the method of level spectroscopy. Unfortunately, this method
becomes unreliable for $J\approx 1/4$ and
$\Delta_1=\Delta_2\approx 1$ because of strong finite-size
effects.

In real chain compounds the exchange interactions are anisotropic.
The microscopic origin of these interactions is the spin-orbit
coupling. The indication on the anisotropy is a dependence of the
saturation field on the direction of the external magnetic field
\cite{Drechsler1}. Though, as a rule, the anisotropy is weak (for
example, for edge-shared cuprate $LiCuVO_4$ ESR detected a $6\%$
anisotropy \cite{prb134445}), it can change the transition point
from commensurate to incommensurate states as well as the behavior
of the model (\ref{H}) in the vicinity of the transition point.
Besides, the frustration parameter $|J_2/J_1|=J$ estimated for
some edge-sharing copper oxides is close to the quantum critical
point $1/4$ (for example, $J\sim 0.28-0.3$ for compound
$Li_2ZrCuO_4$ \cite{PRL07}). Therefore, taking into account both
the frustration effects and the small exchange anisotropy near the
transition point can be important for the analysis of the
experimental data related to these compounds.

In the isotropic case of (\ref{H}) the ferromagnetic state is
($N+1$)-fold degenerated at $0<J<1/4$. Weak easy-plane anisotropy
$\Delta_1,\Delta_2<1$ lifts this degeneracy and the ground state
is in the sector with total $S^z=0$ at small $J$. One can expect
that the increase of $J$ induces the phase transition at some
$J_c$ to the incommensurate phase with $S^z=0$. Besides, the
character of this transition can be different from that in the
isotropic case.

In our analysis we focus on the behavior of the F-AF model
(\ref{H}) near the transition point from the commensurate to the
incommensurate ground state and the influence of the weak
anisotropic interaction on the $T=0$ phase diagram. For simplicity
we concentrate our attention on the particular case of the
Hamiltonian (\ref{H}) with $\Delta_2=1$
\begin{equation}
H=-\sum (S_{n}^{x}S_{n+1}^{x}+S_{n}^{y}S_{n+1}^{y}+\Delta
S_{n}^{z}S_{n+1}^{z}-\frac{1}{4}) + J\sum (\mathbf{S}_{n}\cdot
\mathbf{S}_{n+2}-\frac{1}{4})  \label{HH}
\end{equation}
(We added here constants for convenience.)

However, we will show that the results for the model (\ref{H})
with both $\Delta_1\neq 1$ and $\Delta_2\neq 1$ are qualitatively
similar to those for the model (\ref{HH}).

The paper is organized as follows. In Sec.II we consider a
qualitative physical picture of the ground state phase diagram of
the model (\ref{HH}) based on the classical approximation. In
Sec.III we study the phase diagram of the model (\ref{HH}) using
the variational mean-field approach. The scaling estimates of the
perturbation theory for the easy-plane case of the model (\ref
{HH}) at $J<1/4$ are presented in Sec.IV. In Sec.V we estimate
infrared divergencies in the perturbation theory near the
transition point $J=1/4$. Sec.VI is devoted to the phase
transition in the easy-axis case of the model (\ref{HH}). In
Sec.VII we present the phase diagram of the model (\ref{H}) in the
case $\Delta_1=\Delta_2$ and summarize our results.

\section{Classical approximation}

Let us start from the classical picture of the ground state of the
model (\ref{HH}). In the classical approximation the spins are
vectors which form the spiral structure with a pitch angle
$\varphi $ between neighboring spins and canted angle $\theta $
\begin{eqnarray}
S_{n}^{x} &=&\frac{1}{2}\cos (\varphi n)\sin \theta  \nonumber \\
S_{n}^{y} &=&\frac{1}{2}\sin (\varphi n)\sin \theta  \nonumber \\
S_{n}^{z} &=&\frac{1}{2}\cos \theta  \label{classrotate}
\end{eqnarray}

The classical energy per site is
\begin{equation}
\frac{E_{\mathrm{cl}}(\varphi ,\theta )}{N}=\frac{1-\Delta }{4}+\frac{\sin
^{2}\theta }{4}\left[ \Delta -\cos \varphi -J(1-\cos (2\varphi ))\right]
\label{Ecl}
\end{equation}
\qquad \qquad

The minimization of the energy (\ref{Ecl}) over the angles
$\varphi$ and $\theta$ shows that there are three regions in ($J$,
$\Delta$) having different classical energies. In the region I
($J<\frac{1}{4}$, $\Delta <1$) the energy is minimized by the
choice of the angles $\varphi =0$ and $\theta = \frac{\pi }{2}$.
These angles correspond to the spin configuration with all spins
pointing along the $x$-axis and the energy is
\begin{equation}
E_{\mathrm{cl,xy}}=0  \label{Eclxy}
\end{equation}

In the region II ($J<\frac{1}{4}$, $\Delta >1$) and
($J>\frac{1}{4}$, $ \Delta -1>\frac{2}{J}\left(
J-\frac{1}{4}\right) ^{2}$) the minimum of the energy is given by
the angle $\theta =0$ (and arbitrary $\varphi $). This is the
fully polarized state with all spins up (or down) and the energy
\begin{equation}
E_{\mathrm{cl,z}}=-N\frac{\Delta -1}{4}  \label{Eclz}
\end{equation}

In the region III ($J>\frac{1}{4}$, $\Delta -1<\frac{2}{J}\left(
J-\frac{1}{4}\right) ^{2}$) the classical approximation shows
helical spin structure in the $x$-$y$ plane. The corresponding
angles are
\begin{eqnarray}
\varphi &=&\cos ^{-1}\frac{1}{4J}  \nonumber \\
\theta &=&\frac{\pi }{2}  \label{phicl}
\end{eqnarray}
and the classical ground state energy is
\begin{equation}
E_{\mathrm{cl,sp}}=-\frac{N}{2J}\left( J-\frac{1}{4}\right) ^{2}
\label{Eclspiral}
\end{equation}

The phase boundaries in the classical approximation for the model
(\ref{HH}) are shown in Fig.1 by thin dashed lines. One can see
from Fig.1 that the transition between phases I and II takes place
on the isotropic line $\Delta=1$. This transition is a simple
spin-flop, which is certainly of the first-order type. In the
easy-axis case $\Delta>1$ the increase of the NNN exchange $J$
leads to the first-order transition to the helical phase III with
finite value of $\varphi$ (\ref{phicl}), which is
$\varphi=(8(\Delta-1))^{1/4}$ at $\Delta\to 1$. In contrast to the
easy-axis case, in the easy-plane part of the phase diagram the
transition to the helical phase occurs at $J=1/4$, where the pitch
angle $\varphi=0$, indicating the second-order type of this
transition.

\begin{figure*}
\includegraphics{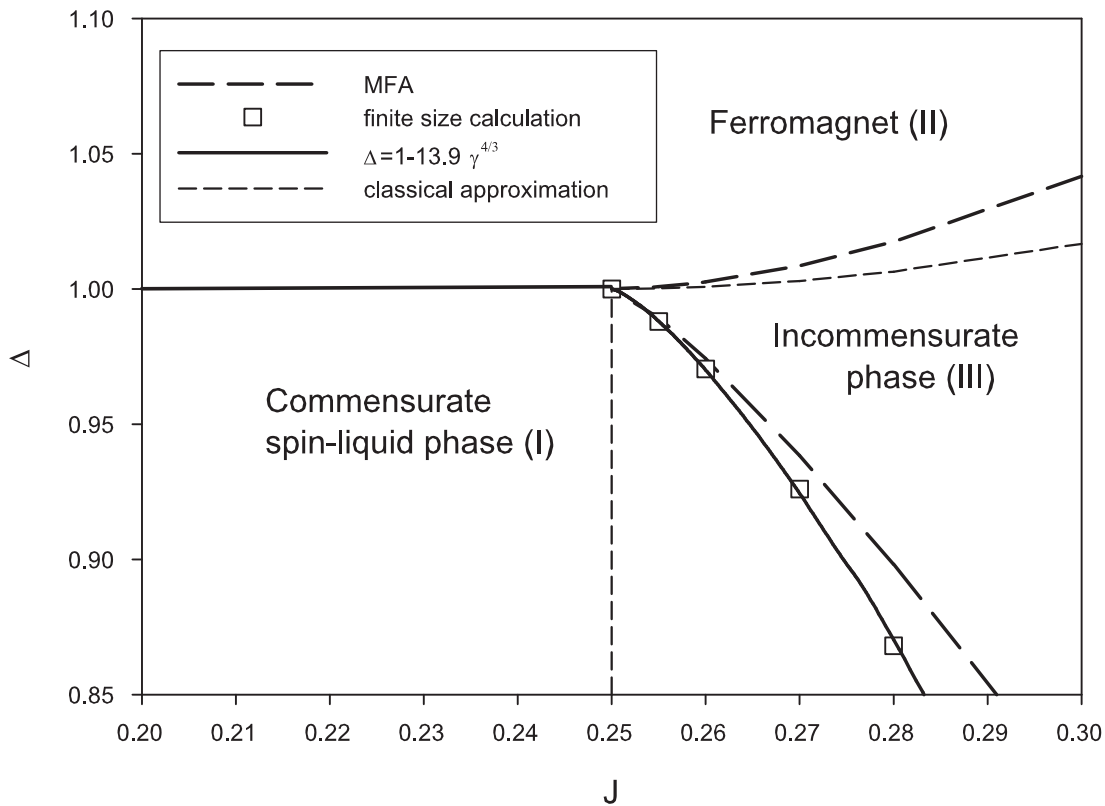}
\caption{The phase diagram of the model (\ref{HH}).} \label{fig_1}
\end{figure*}

\section{Mean field approach}

To study the model (\ref{HH}) we use the variational mean-field
approach (MFA) developed in Ref.\cite{DK04,DK06}. According to
this approach, we follow the classical picture and transform the
local axes on $n$-th site by a rotation about the $Z$ axis by
$\varphi n$ and then by a rotation about the $Y$ axis by $\theta$.
The transformation to new spin-$\frac{1}{2}$ operators
$\mathbf{\eta }_{n}$ has a form
\begin{equation}
\mathbf{S}_{n}=R_{z}(\varphi n)R_{y}(\theta)\mathbf{\eta}_{n}
\label{rotate}
\end{equation}
where $R_{y}(\theta)$ and $R_{z}(\varphi n)$ are the operators of
the corresponding rotations.

The second step is the Jordan-Wigner transformation to the
obtained Hamiltonian in terms of the $\mathbf{\eta }$ operators.
This transformation maps the $\mathbf{\eta }$-spin model onto the
model of interacting spinless fermions, which is then treated by
the mean-field approximation including superconductor like
correlations. The pitch and canted angles $\varphi $ and $\theta $
are variational parameters in this approach. We omit here the
details of this approach, because it is simple modification of
that done in Ref.\cite{DK04,DK06}, and we present here only the
results of this approximation.

Generally, the phase diagram of the model (\ref{HH}) in the MFA
contains the same three phases as predicted by the classical
approximation and the boundaries between the phases are shown in
Fig.1 by thick dashed lines. In the region I (see Fig.1) the MFA
shows the non-zero magnetization in the easy $x-y$ plane. In the
region II the fully polarized state
$\left|\uparrow\uparrow\ldots\uparrow\right\rangle$ represents the
ground state. In the region III the MFA shows helical spin
structure in the $x-y$ plane. However, as can be seen in Fig.1,
the boundary between the phases I and III is substantially
shifted. In the MFA this boundary in the vicinity of the point
($J=\frac{1}{4}$, $\Delta =1$) is approximately given by
\begin{equation}
\alpha\approx 8.05\gamma ^{1.25}  \label{MFAxytrline}
\end{equation}
where $\alpha =1-\Delta $ and $\gamma =J-\frac{1}{4}$.

The boundary between the phases II and III in the MFA is described
by the equation
\begin{equation}
\Delta \approx 1 + 6.3\gamma^{1.7}  \label{MFAztrline}
\end{equation}

Certainly, there is no LRO in the $x$-$y$ plane in the phases I
and III and in this respect the MFA is incorrect. However, the MFA
gives a good estimate for the ground state energy in those phases.
For example, in the phase I at $J=0$ the MFA reproduces correctly
non-trivial critical exponent for the ground state energy
\begin{equation}
\delta E_{0}\approx -0.063N\alpha ^{3/2}  \label{MFAEJ0}
\end{equation}

This estimate differs on 16\% in numerical factor from the exact
result \cite{XXZ}
\begin{equation}
\delta E_0 = -\frac{N\alpha ^{3/2}}{3\sqrt{2}\pi }  \label{EJ0}
\end{equation}

The MFA shows that the critical exponent $3/2$ for the ground
state energy remains up to the point $J=1/4$, where the behavior
of the ground state energy is changed to
\begin{equation}
\delta E_0 \approx -0.07 N \alpha^{9/7}  \label{MFAEJ14}
\end{equation}

As was shown in Ref.\cite{DK06} the MFA gives also a good estimate
for a critical exponent of the ground state energy in the
isotropic case $\Delta =1$ of the helical phase III:
\begin{equation}
\delta E_0 \approx -1.585 N \gamma^{12/7}  \label{MFAEg0}
\end{equation}

As will be shown below, the estimates of the ground state energies
and the phase boundaries in the MFA given by Eqs.(\ref
{MFAxytrline}),(\ref{MFAztrline}) are in a good accordance with
scaling estimates and finite-size calculations.

\section{Perturbation theory for easy-plane case at $J<1/4$}

We are interested in the behavior of the model (\ref{HH}) in the
vicinity of the isotropic case $\Delta =1$. For this aim it is
natural to develop the perturbation theory (PT) in small parameter
$\alpha =1-\Delta $
\begin{eqnarray}
H &=&H_{0}+V_{J}+V_{\delta }  \nonumber \\
H_{0} &=&-\sum (\mathbf{S}_{n}\cdot \mathbf{S}_{n+1}-\frac{1}{4})  \nonumber
\\
V_{J} &=&J\sum (\mathbf{S}_{n}\cdot \mathbf{S}_{n+2}-\frac{1}{4})  \nonumber
\\
V_{\alpha } &=&\alpha \sum S_{n}^{z}S_{n+1}^{z}  \label{PTJ0}
\end{eqnarray}

At first let us consider the most simple case $J=0$, where the
ground-state energy at $\alpha\ll 1$ is given by Eq.(\ref{EJ0}).
The ground state of $H_0$ is ferromagnetic and is degenerated with
respect to total $S^z$. The perturbation $V_{\alpha }$ splits this
degeneracy and in the first order in $\alpha $ we have
\begin{equation}
\left\langle \psi (S^{z})\right| V_{\alpha }\left| \psi
(S^{z})\right\rangle =\frac{\alpha }{4}\frac{(4S^z-N)}{N-1}
\end{equation}

Thus, the first order shows that one should develop the PT for the
lowest state $\left| \psi _{0}\right\rangle $ having total spin
$S=\frac{N}{2}$ and $S^{z}=0$. The perturbation series for the
ground state energy can be written in a form:
\begin{equation}
E_{0}(\alpha )=\left\langle \psi _{0}\right| V_{\alpha }+V_{\alpha
}\frac{1}{E_{0}-H_{0}}V_{\alpha }+\ldots \left| \psi
_{0}\right\rangle  \label{Eseries}
\end{equation}

Suppose that the main contributions to the energy are given by
low-lying excitations, which for isotropic ferromagnet with the
spectrum $\varepsilon _{k}= k^2/2$ behave as
\begin{equation}
E_{k}-E_{0}\sim N^{-2}  \label{dEk2}
\end{equation}

The higher orders of the perturbation series contain more
dangerous denominators and, therefore, possibly have higher powers
of the infrared divergency. Therefore, we use scaling arguments to
estimate the critical exponent for the ground-state energy. Below
we will take care only of powers of divergencies and omit
numerical factors.

Let us assume that the matrix elements of the perturbation
operator $ V_{\alpha }$ between low-lying states $\left| \psi
_{n}\right\rangle $ involved into the PT (having $S^{z}=0$ but
different total $S$) at $ N\to \infty $ behave as
\begin{equation}
\left\langle \psi _{i}\right| V_{\alpha }\left| \psi _{j}\right\rangle \sim
\alpha  \label{Vaexp}
\end{equation}

Collecting the most divergent parts in all orders of the PT, the
correction to the ground state energy takes a form:
\begin{equation}
E_{0}(\alpha )= \left\langle \psi _{0}\right| V_{\alpha }\left|
\psi _{0}\right\rangle \sum_{m=0}^{\infty }c_{m}x^{m} = \alpha
f_{\alpha}(x) \label{Eseries2}
\end{equation}
where $c_{m}$ are unknown constants and
\begin{equation}
x\sim \frac{\left\langle \psi _{i}\right| V_{\alpha }\left| \psi
_{k}\right\rangle }{E_{k}-E_{0}}\sim \alpha N^{2}  \label{xaN2}
\end{equation}
is a scaling parameter, which absorbs the infrared divergencies.

The scaling function $f_{\alpha}(x)$ at $x\to 0$ is given by the
first order correction. In the thermodynamic limit ($x\to \infty
$) the behavior of $f_{\alpha}(x)$ is generally unknown, but the
natural condition $E_{0}(\alpha )\sim N$ at $N\to \infty $
requires
\begin{equation}
f_{\alpha}(x)\sim \sqrt{x}  \label{f(x)}
\end{equation}
and, finally
\begin{equation}
E_{0}(\alpha )\sim -N\alpha ^{3/2}  \label{Eexp}
\end{equation}

The obtained expression is in agreement with the exact equation
(\ref{EJ0}) for the ground state energy, which justifies our
assumption about the behavior of the matrix elements
(\ref{Vaexp}).

Moreover, exploiting the fact that the system in the region
$\Delta <1$ is in a spin-liquid phase, the correction to the
ground-state energy has a form \cite{blote}
\begin{equation}
E_0 = Ne_0 - \frac{\pi cv_{\mathrm{sound}}}{6N}  \label{Exxz}
\end{equation}
where $e_0$ is the ground state energy at $N\to\infty$ and the
central charge is $c=1$ in our case.

In order to reproduce such $1/N$ correction to the energy, the
asymptotic of $f_{\alpha}(x)$ at large $x$ should have a form
\begin{equation}
f_{\alpha}(x)= a\sqrt{x}+\frac{b}{\sqrt{x}}  \label{f(x)corr}
\end{equation}
with some constants $a$ and $b$.

So, from Eq.(\ref{Eseries2}) we find
\begin{equation}
E_{0}(\alpha )= -Na\alpha ^{3/2}-\frac{b\alpha ^{1/2}}{N}
\end{equation}
and, therefore,
\begin{equation}
v_{\mathrm{sound}}\sim \sqrt{\alpha }  \label{va12}
\end{equation}
which agrees with an exact result
$v_{\mathrm{sound}}=\sqrt{\alpha/2}$ at $\alpha\ll 1$
\cite{vsound}. Thus, the scaling estimates give us the correct
exponent for the sound velocity as well.

Now let us consider the PT (\ref{PTJ0}) containing both channels
$V_{\alpha } $ and $V_{J}$. In order to estimate the powers of
divergency of high-orders in this PT one needs to know the
$N$-dependence of the matrix elements $\left\langle \psi
_{i}\right| V_{J}\left| \psi _{j}\right\rangle $ . In general, it
is unknown. However, one can restore these matrix elements from
the known exact expression for NNN spin correlator in the ground
state $\left| \psi_0 (\alpha )\right\rangle$ at $J=0$ and some
small value of $\alpha $ \cite{kato}:
\begin{equation}
\left\langle \psi _{0}(\alpha )\right| ( \mathbf{S}_{n}\cdot
\mathbf{S}_{n+2}- \frac{1}{4} ) \left| \psi _{0}(\alpha
)\right\rangle = -\frac{\sqrt{2}}{3\pi } \alpha ^{3/2}
\label{corr2}
\end{equation}
or, in other words,
\begin{equation}
\left\langle \psi _{0}(\alpha )\right| V_{J}\left| \psi _{0}(\alpha
)\right\rangle =-\frac{\sqrt{2}}{3\pi }\alpha ^{3/2}JN  \label{VJa}
\end{equation}

On the other hand, collecting all contributions of the PT to the
linear term in $J$ similar to done in Eq.(\ref{Eseries2}), we
arrive at a scaling form in small parameter $\alpha $
\begin{equation}
\left\langle \psi _{0}(\alpha )\right| V_{J}\left| \psi
_{0}(\alpha )\right\rangle \sim \left\langle \psi _{i}\right|
V_{J}\left| \psi _{j}\right\rangle \cdot f_J(x) \label{VJa2}
\end{equation}
with $x=\alpha N^2$.

The comparison of Eqs.(\ref{VJa}) and (\ref{VJa2}) immediately
leads to the results
\begin{equation}
\left\langle \psi _{i}\right| V_{J}\left| \psi _{j}\right\rangle \sim JN^{-2}
\label{VJexp}
\end{equation}
and
\begin{equation}
f_J(x)\sim x^{3/2}  \label{f1(x)}
\end{equation}

So, the matrix elements $\left\langle \psi _{i}\right| V_{J}\left|
\psi _{j}\right\rangle $ are small enough to eliminate dangerous
denominators:
\begin{equation}
y\sim \frac{\left\langle \psi _{i}\right| V_{J}\left| \psi _{j}\right\rangle
}{E_{k}-E_{0}}\sim J  \label{yJ}
\end{equation}
which, in turn, implies the absence of infrared divergencies in
$V_J$ channel. Thus, the perturbation $V_{J}$ does not form a
scaling parameter and the ground state energy has regular
expansion in $J$.

It is natural to expect that the behavior of the matrix elements
of the type (\ref{VJexp}) remains the same up to the point
$J=1/4$. It results in the expression for the ground state energy
at $J<1/4$:
\begin{equation}
E_{0}= -N\alpha ^{3/2}g_J(J)  \label{ExyJa}
\end{equation}
where $g_J(J)$ is some unknown smooth function, which at small $J$
has the expansion in accordance with exact results (\ref{EJ0}),
(\ref{VJa}):
\begin{equation}
E_{0}=-\frac{N\alpha ^{3/2}}{3\sqrt{2}\pi }(1+2J)  \label{ExyJ0}
\end{equation}

However, at approaching to the point $J=1/4$ one should take into
account that the excitation spectrum is $\varepsilon _{k}=\left(
\frac{1}{2} -2J\right) k^{2}$ and the excitation energies become
\begin{equation}
E_{k}-E_{0}\sim \frac{\frac{1}{4}-J}{N^{2}}  \label{dEk2J}
\end{equation}

This modifies the scaling parameter
\begin{equation}
x\sim \frac{\alpha N^{2}}{\frac{1}{4}-J}  \label{xaN2J}
\end{equation}
and the expression for the energy
\begin{equation}
E_{0}\sim -\frac{N\delta ^{3/2}}{\sqrt{\frac{1}{4}-J}}
\end{equation}

Similarly, the sound velocity at $J\to 1/4$ behaves as
\begin{equation}
v_{\mathrm{sound}}\sim \sqrt{\frac{1}{4}-J}\sqrt{\alpha }  \label{vag12}
\end{equation}

\section{Perturbation theory near the transition point $J=1/4,\Delta =1$}

At $J=1/4$ and $\Delta =1$ the ferromagnetic ground state becomes
degenerated with a singlet spiral state \cite{Hamada}. At $\Delta
<1$ the ground state obviously lies in the $S^z=0$ sector.
Therefore, in order to determine the transition line between the
phases I and III one should develop the PT both to the
ferromagnetic state with $S^z=0$ and to the singlet spiral state.

\subsection{The PT to the ferromagnetic state with $S^{z}=0$}

Let us represent the Hamiltonian in a form
\begin{eqnarray}
H &=&H_{0}+V_{\delta }+V_{\gamma }  \nonumber \\
H_{0} &=&-\sum (\mathbf{S}_{n}\cdot \mathbf{S}_{n+1}-\frac{1}{4})+\frac{1}{4}%
\sum (\mathbf{S}_{n}\cdot \mathbf{S}_{n+2}-\frac{1}{4})  \nonumber \\
V_{\alpha } &=&\alpha \sum S_{n}^{z}S_{n+1}^{z}  \nonumber \\
V_{\gamma } &=&\gamma \sum (\mathbf{S}_{n}\cdot \mathbf{S}_{n+2}-\frac{1}{4})
\label{PTJ14}
\end{eqnarray}

We assume that the behavior of\ the matrix elements of operators
$V_{\alpha } $ and $V_{\gamma }$ remains the same as in the region
$J<1/4$ (see Eq.(\ref{Vaexp}) and Eq.(\ref{VJexp})). However, the
scaling parameters are modified due to the changing in
one-particle excitation spectrum, which is $\varepsilon
_{k}=k^{4}/8$ at $J=1/4$. So, the low-lying excited states
involved in the PT\ (\ref{PTJ14}) behave as
\begin{equation}
E_{k}-E_{0}\sim N^{-4}  \label{dEk4}
\end{equation}

Now according to Eqs.(\ref{xaN2}), (\ref{yJ}) both channels
$V_{\alpha }$ and $V_{\gamma }$ produce the scaling parameters:
\begin{eqnarray}
x &=&\alpha N^4  \nonumber \\
y &=&\gamma N^2  \label{xyJ14}
\end{eqnarray}

Thus, as follows from Eq.(\ref{Eseries2}) exactly at $J=1/4$
($y=0$) the ground state energy can be written in a scaling form
\begin{equation}
E_{0}(\alpha )= -N\alpha^{5/4}f(x)  \label{Ea54}
\end{equation}

This scaling and the critical exponent is confirmed by numerical
calculations, where the function $f(x)$ is calculated on finite
chains with different $N$ and $\alpha$ for the ground state with
$k=0$ (see Fig.2). As one can see on Fig.2, all data lie perfectly
on one curve $f(x)$ and in the thermodynamic limit the function
$f(x)\to 0.08$. We show in Fig.2 that the same scaling
(\ref{Ea54}) is valid for the lowest excited state with $k=\pi$ as
well and that the corresponding scaling function has the same
thermodynamic limit $f(x)\to 0.08$.

\begin{figure*}
\includegraphics{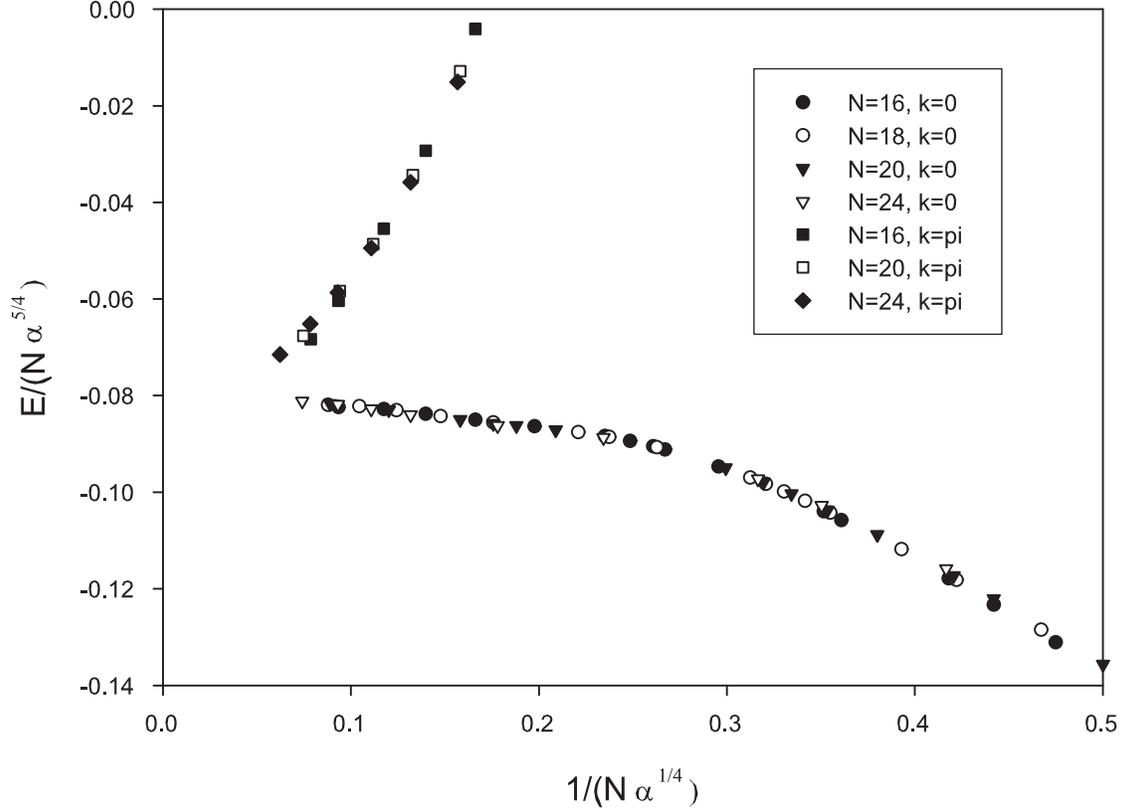}
\caption{The scaling function $f(x)$ in Eq.(\ref{Ea54}) for the
ground state energy and the lowest excited state at $J=1/4$.}
\label{fig_2}
\end{figure*}

The system at $J=1/4$ and $\alpha >0$ is in a spin-liquid phase,
which is verified by $1/N$ behavior of low-lying excitations
calculated on finite chains (see Fig.3). From the scaling equation
(\ref{Ea54}) we can extract also the critical exponent for the
sound velocity:
\begin{equation}
v_{\mathrm{sound}}\sim \alpha ^{3/4}  \label{va34}
\end{equation}

\begin{figure*}
\includegraphics{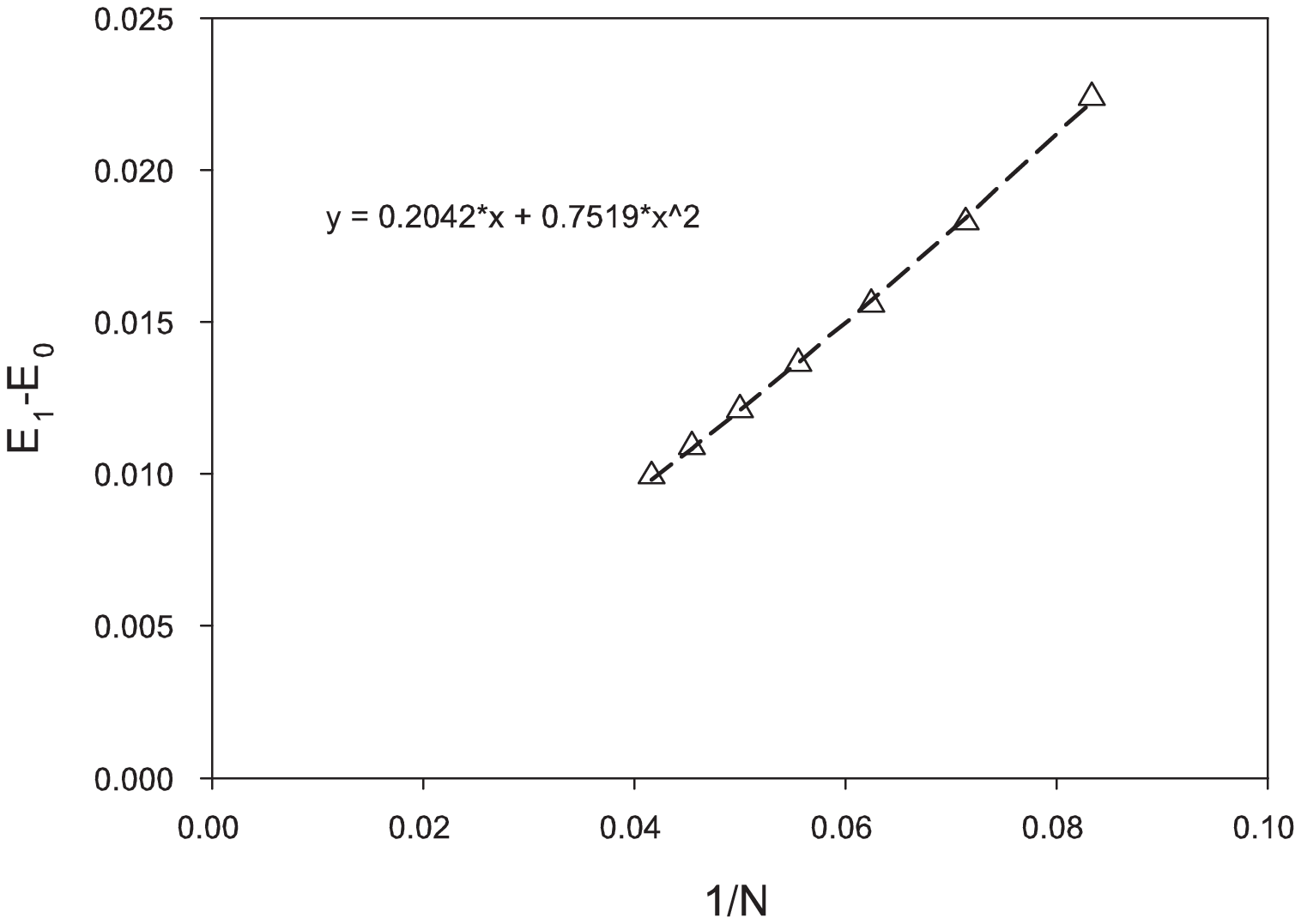}
\caption{$N$-dependence of the energy gap between the ground state
and the lowest excitation of the model (\ref{HH}) at $J=1/4$ and
$\Delta=0.96$.} \label{fig_3}
\end{figure*}

In case when both $V_{\alpha }$ and $V_{\gamma }$ play simultaneously, the
scaling estimates (\ref{xyJ14}) give
\begin{equation}
E_{0}(\alpha ,\gamma )= -N\alpha ^{5/4}f(x,y)  \label{Eag54}
\end{equation}

In the thermodynamic limit, when both $x\to \infty $ and $y\to
\infty $, the scaling function $f(x,y)$ becomes a function of one
variable (independent on $N$)
\begin{equation}
\nu =\frac{y^{2}}{x}=\frac{\gamma ^{2}}{\alpha }  \label{z}
\end{equation}
and the ground state energy takes a form
\begin{equation}
E_{0}(\alpha ,\gamma )= -N \alpha^{5/4} g(\nu )  \label{Ez}
\end{equation}


\subsection{The PT to the singlet spiral state}

The PT to the singlet spiral state with pitch angle $\varphi$ in
the isotropic case $\alpha =0$ was developed in Ref.\cite{DKR},
where it was found that the energy has a scaling form
\begin{equation}
E(0,\gamma ,\varphi )=-N\frac{\gamma \varphi ^{2}}{2} +
N\varphi^{5}f(\gamma N^{2},\varphi N)  \label{Egphi}
\end{equation}
where the first term comes from the first order of the PT in
$\gamma $ and the second one originates from the scaling estimates
of the infrared divergencies of higher-orders of the PT.

Comparison of Eq.(\ref{Eag54}) and Eq.(\ref{Egphi}) leads to a
general expression for the energy, which correctly reduces to both
cases at $\varphi \to 0$ and $\alpha \to 0$
\begin{equation}
E(\alpha ,\gamma ,\varphi) = -N\frac{\gamma \varphi ^{2}}{2} +
N\varphi^5 f(\alpha N^4,\gamma N^2,\varphi N) \label{Eagphi}
\end{equation}
(in fact, this equation can be derived in a similar manner as was
done in Ref.\cite{DKR})

In the thermodynamic limit, when all variables in the scaling
function in Eq.(\ref{Eagphi}) tends to infinity, the scaling
dependence transforms to a function of two variables
\begin{equation}
E(\alpha ,\gamma ,\varphi ) = -N\frac{\gamma \varphi ^{2}}{2} + N
\varphi^5 g(\mu ,\nu )  \label{Emunu}
\end{equation}
where
\begin{eqnarray}
\mu &=&\frac{\alpha }{\varphi ^{4}}  \nonumber \\
\nu &=&\frac{\gamma ^{2}}{\alpha }  \label{munu}
\end{eqnarray}

Generally, the function $g(\mu ,\nu )$ is unknown. However, we can
identify some of its properties. At first, in the limit $\varphi
\to 0$ we should reproduce Eq.(\ref{Ez}). Moreover, in the
spin-liquid phase the spiral states with $\varphi \sim N^{-1}$
should describe sound-like excitations with the sound velocity
(\ref{va34}). These requirements suggest that in the limit $\mu
\to \infty $ the function $g(\mu ,\nu )$ has an asymptotic
behavior
\begin{equation}
\lim_{\mu \rightarrow \infty }g(\mu ,\nu )\sim -\mu
^{5/4}g_{1}(\nu )+\mu ^{3/4}g_{2}(\nu )+o(\mu ^{3/4})
\label{muinf}
\end{equation}

One can check, that this expression reproduces the sound-like
excitations at $\gamma =0$:
\begin{equation}
E(\alpha ,0,\varphi )-E(\alpha ,0,0)\sim N\alpha ^{3/4}\varphi ^{2}\sim
\frac{\alpha ^{3/4}}{N}  \label{phi2}
\end{equation}

In the limit $\mu \to 0$, according to Ref.\cite{DKR} we have
\begin{equation}
\lim_{\mu \to 0}g(\mu ,\nu )\sim A+o(1)  \label{mu0}
\end{equation}
where constant $A$ describes the excitation spectrum at the
transition point $\alpha =0$ and $\gamma =0$. Finite-size
calculations give for this constant the value $A\approx 0.0065$.

Summarizing all above we extract explicitly the corresponding
terms and obtain the following expression:
\begin{equation}
\frac{1}{N}E(\alpha ,\gamma ,\varphi ) = -\frac{\gamma \varphi
^{2}}{2} - \alpha ^{5/4}g_{1}(\nu ) + \alpha ^{3/4}\varphi
^{2}g_{2}(\nu ) + A\varphi ^{5} + \varphi ^{5}g_3(\mu ,\nu )
\label{Emunu2}
\end{equation}
where the function $g_3(\mu ,\nu )$ has limits
\begin{eqnarray}
\lim_{\mu \to 0}g_3(\mu ,\nu ) &\sim &o(1)  \nonumber \\
\lim_{\mu \to \infty }g_3(\mu ,\nu ) &\sim &o(\mu ^{3/4})
\label{gmu}
\end{eqnarray}

The minimization of the energy (\ref{Emunu2}) over the pitch angle
$\varphi $ gives equation for $\varphi _{\min }$
\begin{equation}
\frac{\partial E(\alpha ,\gamma ,\varphi )}{\partial \varphi }=0
\label{dEdphi}
\end{equation}
or after some algebra
\begin{equation}
\gamma =\varphi ^{3}\left[ 5A+2\mu ^{3/4}g_{2}(\nu )+5g_3(\mu ,\nu
)-4\mu \frac{\partial g_3(\mu ,\nu )}{\partial \mu }\right]
\label{Eqphimin}
\end{equation}

We see that in the isotropic limit $\alpha \ll \varphi ^{4}$ ($\mu
\to 0$) the pitch angle is defined by the constant term in
right-hand side of Eq.(\ref{Eqphimin}) (using Eq.(\ref{gmu}))
\begin{equation}
\varphi _{\min }=\left( \frac{\gamma }{5A}\right) ^{1/3}
\end{equation}
which reproduces the result of Ref.\cite{DKR}.

In order to find the commensurate-incommensurate transition line,
where the pitch angle $\varphi_{\min }$ vanish, it is more
convenient to rewrite Eq.(\ref{Eqphimin}) in a form:
\begin{equation}
\frac{\gamma }{\alpha ^{3/4}} - 2g_{2}(\nu ) = \frac{1}{\mu
^{3/4}}\left[ 5A+5g_3(\mu ,\nu )-4\mu \frac{\partial g_3(\mu ,\nu
)}{\partial \mu }\right] \label{Eqphimin2}
\end{equation}

From Eq.(\ref{Eqphimin2}) and Eq.(\ref{gmu}) one can see that the
right-hand side of Eq.(\ref{Eqphimin2}) tends to zero at $\mu \to
\infty$, which corresponds to the limit $\varphi\to 0$. The
left-hand side of (\ref{Eqphimin2}) is independent on $\varphi$
and vanishes on the transition line:
\begin{equation}
\gamma = 2g_{2}(0)\alpha ^{3/4}  \label{trline}
\end{equation}
(we note, that on the transition line (\ref{trline}) $\nu =0$).

Hence, at approaching to the transition line (\ref{trline}) the
pitch angle $\varphi _{\min }$ smoothly goes to zero. So, the line
(\ref{trline}) determines the second-order transition line between
the commensurate spin-liquid phase I with $\varphi =0$ and the
incommensurate spiral phase III with $\varphi \neq 0$.

Another question that can be studied concerns the low-lying
excitations in the incommensurate phase. According to
Eq.(\ref{dEdphi}) the behavior of the energy near $\varphi_{\min}$
is expanded as:
\begin{equation}
E(\alpha ,\gamma ,\varphi )=E(\alpha ,\gamma ,\varphi _{\min })+\frac{\left(
\varphi -\varphi _{\min }\right) ^{2}}{2}\frac{\partial ^{2}E(\alpha ,\gamma
,\varphi )}{\partial \varphi ^{2}}  \label{Eexpans}
\end{equation}

The second-order derivative of the energy at
$\varphi=\varphi_{\min}$ can be estimated as
\begin{equation}
\frac{\partial^2 E(\alpha ,\gamma ,\varphi _{\min })}{\partial
\varphi^2} \sim N\gamma  \label{d2Edphi2}
\end{equation}

Thus, the states with
\begin{equation}
\varphi _{k}=\varphi _{\min }\pm \frac{2\pi }{N}k  \label{phishift}
\end{equation}
describe gapless excitations with the energy
\begin{equation}
\delta E\sim \frac{\gamma }{N}  \label{dEspiral}
\end{equation}

Certainly, there is no helical LRO in the spiral phase and the
spin correlations decay on large distances. However, the nature of
the spiral phase manifest itself in the incommensurate position
$q_{\max }$ of the maximum of structure factor
\begin{equation}
S(q)=\sum_{n,r}e^{iqr}\left\langle \mathbf{S}_{n}\cdot \mathbf{S}
_{n+r}\right\rangle   \label{Sq}
\end{equation}

When the O(3) rotation symmetry is broken by the anisotropic term
$V_{\alpha}$, the incommensurate nature of the spiral phase
remains in the $x$-$y$ plane. So, in this case we associate the
pitch angle of the spiral $\varphi$ with the position of maximum
of the structure factor $q_{\max}$ in the easy plane
\begin{equation}
S^{xx}(q)=\sum_{n,r}e^{iqr}\left\langle
S_{n}^{x}S_{n+r}^{x}+S_{n}^{y}S_{n+r}^{y}\right\rangle   \label{Sxxq}
\end{equation}

The numerical calculations on finite chains show that for a fixed
small value of $\gamma$, $q_{\max}$ decreases via consecutive
sharp jumps on the value $\frac{2\pi}{N}$ from some finite value
at $\alpha=0$ to zero on the transition line
\begin{equation}
\alpha\approx 13.9\gamma^{4/3}  \label{trlinenum}
\end{equation}
where the incommensurate phase III terminates (see Fig.1) and the
transition into commensurate spin-liquid phase takes place. Thus,
the numerical calculation confirms the found critical exponent for
the transition line (\ref{trline}). The fact that the pitch angle
$\varphi$ tends to zero at approaching to the transition line
ensures that this line is the second-order transition.

\section{The easy-axis case}

In the easy axis case for $J<1/4$, the fully polarized state
$\left| \uparrow \uparrow \ldots \uparrow \right\rangle $ is
evidently the ground state. In the region $J>1/4$ one should
compare the fully polarized state energy with the energy of the
spiral state. The finite-size numerical calculations show that for
a fixed small $\gamma$ the increasing of easy-axis anisotropy
leads to the decrease of the pitch angle $\varphi $, but the
ground state remains in the sector with total $S^z=0$. At some
critical value of $\Delta_c$ the transition from the state with
$S^z=0$ and some finite value of the pitch angle $\varphi$ to the
fully polarized state occurs. Thus, in contrast to the easy-plane
case, the transition from the spiral phase to the ferromagnetic
phase is the first-order one.

The finite-size numerical calculations also show that in the
spiral region $1<\Delta <\Delta_c$, it is sufficient to take into
account only the first-order correction in $(\Delta -1)$ to the
spiral state. That is the energy of the spiral state is
\begin{equation}
E_{sp}=-aN\gamma ^{5/3}-N\frac{\Delta -1}{12}  \label{Espea}
\end{equation}
and the transition to the fully polarized state with the energy
\begin{equation}
E_{f}=-N\frac{\Delta -1}{4}  \label{Ef}
\end{equation}
takes place at
\begin{equation}
\Delta _{c}=1+6a\gamma^{5/3}  \label{Deltac}
\end{equation}

Unfortunately, the finite-size calculations do not allowed to find
the factor $a$ in Eq.(\ref{Deltac}) because of irregular behavior
of $\Delta_c$ with $N$. However, we believe that the MFA gives a
good estimate for this transition line (\ref{MFAztrline}).

\begin{figure*}
\includegraphics{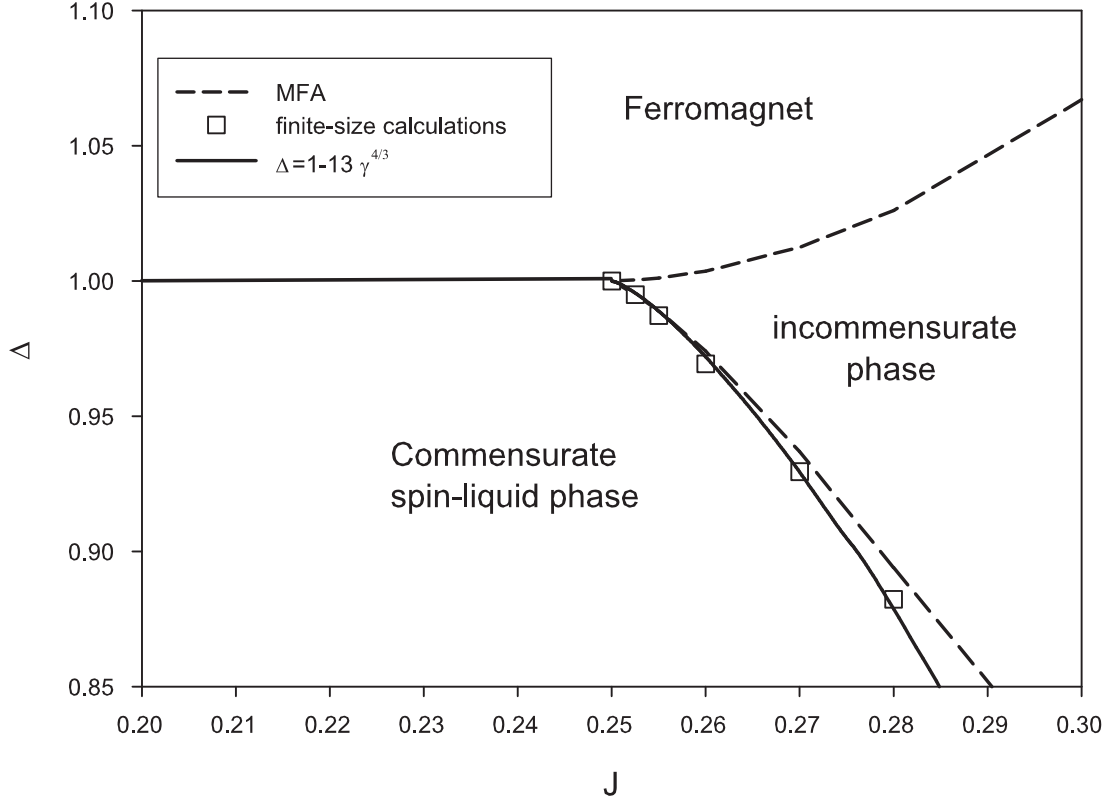}
\caption{The phase diagram of the model (\ref{H}) with
$\Delta_1=\Delta_2=\Delta$.} \label{fig_4}
\end{figure*}

\section{Summary}

We have studied spin-1/2 zigzag chain with weakly anisotropic
ferromagnetic nearest-neighbor and antiferromagnetic
next-nearest-neighbor interactions. It was shown that the ground
state phase diagram consists of three phases: the fully polarized
ferromagnetic phase, the commensurate spin-liquid phase and the
incommensurate phase. Thus, the incommensurate phase established
for the isotropic case survives weak anisotropy of interactions,
though in this case the incommensurate nature of the ground state
reveals itself in the $x-y$ plane.

Using scaling estimates of the infrared divergencies in the
perturbation theory we obtained the scaling expression for the
ground state energy both for commensurate and incommensurate
phases. This allowed us to determine non-trivial critical
exponents in the behavior of the phase transition lines, which
were confirmed by finite-size calculations. We found also that in
the easy-plane case the transition from the commensurate
spin-liquid to the incommensurate phase is of the second order
one, while in the easy-axis case the transition from the fully
polarized state with $S^z=S^z_{max}$ to the incommensurate state
with $S^z=0$ is evidently of the first order.

In this paper we have focused on studying of the model (\ref{HH}),
which is a particular case of more general model (\ref{H}).
However, the obtained results for the model (\ref{HH}) remain
valid at least qualitatively for the model (\ref{H}). As an
example we present in Fig.4 the phase diagram near the transition
point $J=1/4$ of the model (\ref{H}) in the case
$\Delta_1=\Delta_2$. We see that the phase diagram in this case is
very similar even quantitatively to that shown in Fig.1.

\begin{acknowledgments}
We would like to thank S.-L.Drechsler and D.Baeriswyl for valuable
comments related to this work. D.D. thanks the University of
Fribourg for kind hospitality. D.D. was supported by INTAS YS
Grant Nr. 05– 109–4916. The numerical calculations were carried
out with use of the ALPS libraries \cite{alps}.
\end{acknowledgments}

\end{document}